\documentclass[a4paper, 10pt, conference]{ieeeconf}      

\IEEEoverridecommandlockouts                              

\usepackage{amsmath}
\usepackage{amssymb}
\usepackage{ifthen}
\usepackage{color,soul}
\usepackage{txfonts}
\usepackage[linesnumbered]{algorithm2e}
\usepackage{fullpage}
\usepackage{url}
\usepackage{tikz}
\usetikzlibrary{arrows,backgrounds,decorations,decorations.pathmorphing,
	positioning,fit,automata,shapes,snakes,patterns,plotmarks,calc,trees}
\usepackage{subfig}
\usepackage{latexsym}	
\usepackage{xspace}
\usepackage{alltt}
\usepackage[obeyFinal]{todonotes}
\usepackage{wrapfig}
\usepackage[per-mode=symbol]{siunitx}
\usepackage{paralist}
\usepackage{color,colortbl}
\usepackage{pgfgantt}

\usepackage{epsfig,color}
\usepackage{algpseudocode}
\usepackage{mdwlist}
\usepackage{setspace} 
\usepackage{hyperref}
\usepackage[depth=1]{bookmark}
\usepackage{listings}
\usepackage{caption}
\usepackage{xspace}
\usepackage{tikz}
\usepackage{stmaryrd}
\usepackage{multirow}

\usepackage{ltl}
\usepackage{ulem} 
\usepackage{habbas-macros}
\graphicspath{{figs/}}

\newcommand{\Fbl}{Fly-by-Logic}
\newcommand{\FFbl}{FairFly}
\newcommand{\slength}{\ell}
\newcommand{\vlength}{\vec{\slength}}
\newcommand{\robfg}{\rob_{\formula_g}}
\newcommand{\lhn}[1]{\underline{\slength_{#1}}}
\newcommand{\uhn}[1]{\overline{\slength_{#1}}}

\title{\LARGE \bf
\FFbl: A Fair Motion Planner for Fleets of Autonomous UAVs in Urban Airspace
}

\author{Connor Kurtz$^{1}$ and Houssam Abbas$^{1}$
\thanks{$^{1}$The authors are with the School of Electrical Engineering and Computer Science, Oregon State University, OR, USA
        {\tt\small \{kurtzco,houssam.abbas\}@oregonstate.edu}}%
}

\usepackage[a4paper, left=19.1mm, right=13.1mm, bottom=19.1mm, top=43mm]{geometry}
\usepackage{afterpage}

\begin{document}

\maketitle
\thispagestyle{empty}
\pagestyle{empty}

\begin{abstract}
We present a solution to the problem of fairly planning a fleet of Unmanned Aerial Vehicles (UAVs) that have different missions and operators, such that no one operator unfairly gets to finish its missions early at the expense of others - unless this was explicitly negotiated.
When hundreds of UAVs share an urban airspace, the relevant authorities should allocate corridors to them such that they complete their missions, but no one vehicle is accidentally given an exceptionally fast path at the expense of another, which is thus forced to wait and waste energy.
Our solution, FairFly, addresses the fair planning question for general autonomous systems, including UAV fleets, subject to complex missions typical of urban applications.
FairFly formalizes each mission in temporal logic.
An offline search finds the fairest paths that satisfy the missions and can be flown by the UAVs, leading to lighter online control load.
It allows explicit negotiation between UAVs to enable imbalanced path durations if desired.
We present \yhl{three} fairness notions, including one that reduces energy consumption.
We validate our results in simulation, and demonstrate a lighter computational load and less UAV energy consumption as a result of flying fair trajectories.
\end{abstract}

\section{Introduction: What is Fair Usage of Airspace?}
\label{sec:intro}

The growth in Unmanned Aerial Vehicles (UAVs) research is driven by many stakeholders, from different industries and government agencies:
to list a few, 
retailers want to use UAVs to deliver goods faster and with less energy expenditure, 
engineering companies use UAVs to inspect urban infrastructure like rails and solar panels in a more timely manner,
city government can use UAVs for traffic analysis over a wider scale, 
and communications companies envision the creation of `Cellular Networks on Demand' using UAVs as wireless hotspots at times of greater-than-usual demand or in disasters that reduce the fixed network's capacity.
The website Aviation Planning lists over 400 uses of UAVs as of the end of 2018.

The fragmentation in the use cases of UAVs noted above, however, points to a serious challenge facing regulators who want to safely enable advanced urban UAV applications.
Namely, UAVs sharing the same airspace have different operators who respond to different priorities. 
Generally, even if the deadlines for the different missions have been agreed, an operator would still prefer completing its mission earlier rather than later - finishing early can mean higher usage of the UAV, less energy wasted in holding patterns, etc. 
However, this can conflict with \textit{another} operator completing \textit{its} mission as soon as it could. 
Intuitively, if two missions have the same deadline of 10mins, it would be `unfair' to allocate a 1-min motion path to the first drone, and a 9-mins motion path to the second, assuming a more balanced solution exists.
Thus, there remains a need to \textit{balance} the flight durations of all the UAVs, such that all UAVs accomplish their mission within the deadline, and no UAVs are treated unfairly.

This is different from the task of scheduling commercial airliners: in UAV Traffic Management, the number of UAVs sharing a small space is significantly larger, their missions are more complex, their dynamics are more agile, and they are more susceptible to disturbances.
These characteristics require flexible yet robust controllers, which complicates the fairness question beyond a scheduling problem.
Without a \textit{transparent and explicit} fairness mechanism, smaller operators \yhl{are} discouraged from leveraging UAV technology, innovation \yhl{can} be stifled, and the economic benefits of un-piloted aerial systems \yhl{are} foregone.

\yhl{\textbf{Contributions of this work.}}
\label{sec:contributions}
This paper addresses the question of fair and safe motion planning for heterogeneous groups of UAVs which happen to share the same relatively small airspace.
\\ \noindent
1) We provide a computational formulation for the problem of fair allocation of airspace volume in UAV Traffic Management (UTM). This formulation allows complex missions typical of small UAV applications over urban airspace.
\\ \noindent
2) We define 3 notions of fairness, including a notion that 
gives priority to privileged operators. 
Our framework can accommodate other fairness notions.
\\ \noindent
3) We provide an algorithm for solving the fair motion planning problem and demonstrate on quadrotor simulations using controllers that have been shown to work on real-life quadrotors.
\\ \noindent
4) The FairFly framework contributes towards a concrete implementation of equitable access emphasized in the FAA UTM Concept of Operations. 

The paper is organized as follows: Section~\ref{sec:prelims} gives technical preliminaries and presents \Fbl, on top of which we build our solution.
Section~\ref{sec:fair ctrl} presents the Fair Control problem and our solution to it, 
and Section~\ref{sec:exp} presents experimental validation.

\afterpage{\newgeometry{a4paper, left=19.1mm, right=13.1mm, bottom=19.1mm, top=37mm}}

\section{Technical Preliminaries}
\label{sec:prelims}
\textbf{Notation.}
The set of non-negative integers is $\Ne$.
Given a set $X$ and integer $n$, $X^n$ is its $n$-fold Cartesian product, and $\sttraj^n$ is an element of $X^n$, i.e., a sequence of $n$ values from $X$.


\textbf{System model.}
Consider a fleet of $D$ UAVs.
The $n^{th}$ UAV is modeled as a discrete-time dynamical system $x_n[k+1] = f_n(x_n[k],u_n[k]),  x_n[0]\in I_0 \subset \Re^{d}$ \yhl{where $d$ is the dimensionality of the system state.}
The control input applied to the UAV at time $k$ is $u_n[k] \in U \subset \Re^{m}$ \yhl{where $m$ is the dimensionality of the input}.
$I_0$ denotes the set of possible take-off positions, velocities, accelerations, etc.
By concatenating all $D$ states together into a \textit{system state} $x = [x_1, x_2,\ldots, x_D]$ and all inputs into a system input $u=[u_1,\ldots,u_D]$ we get the fleet dynamical system:
\begin{equation}
\label{eq:cont sys}
x[k+1] = f(x[k],u[k]), \quad x[0]\in X_0 \subset \Re^{d\cdot D} 
\end{equation}
Here,  $X_0 = I_0^D$.
Given an initial state $x[0]$ and an input sequence $\inpSig^{H-1} = (u[0],\ldots,u[H-2])$,
the corresponding \textit{trajectory} is the sequence $\sttraj^H = (x[0],\ldots,x[H-1])$ of states that satisfy \eqref{eq:cont sys}.
We will sometimes write it as $\sttraj(\inpSig^{H-1})$.
Our method applies to nonlinear dynamical systems in general, not only UAVs.
\begin{figure}[t]
\centering
\includegraphics[width=0.7\linewidth]{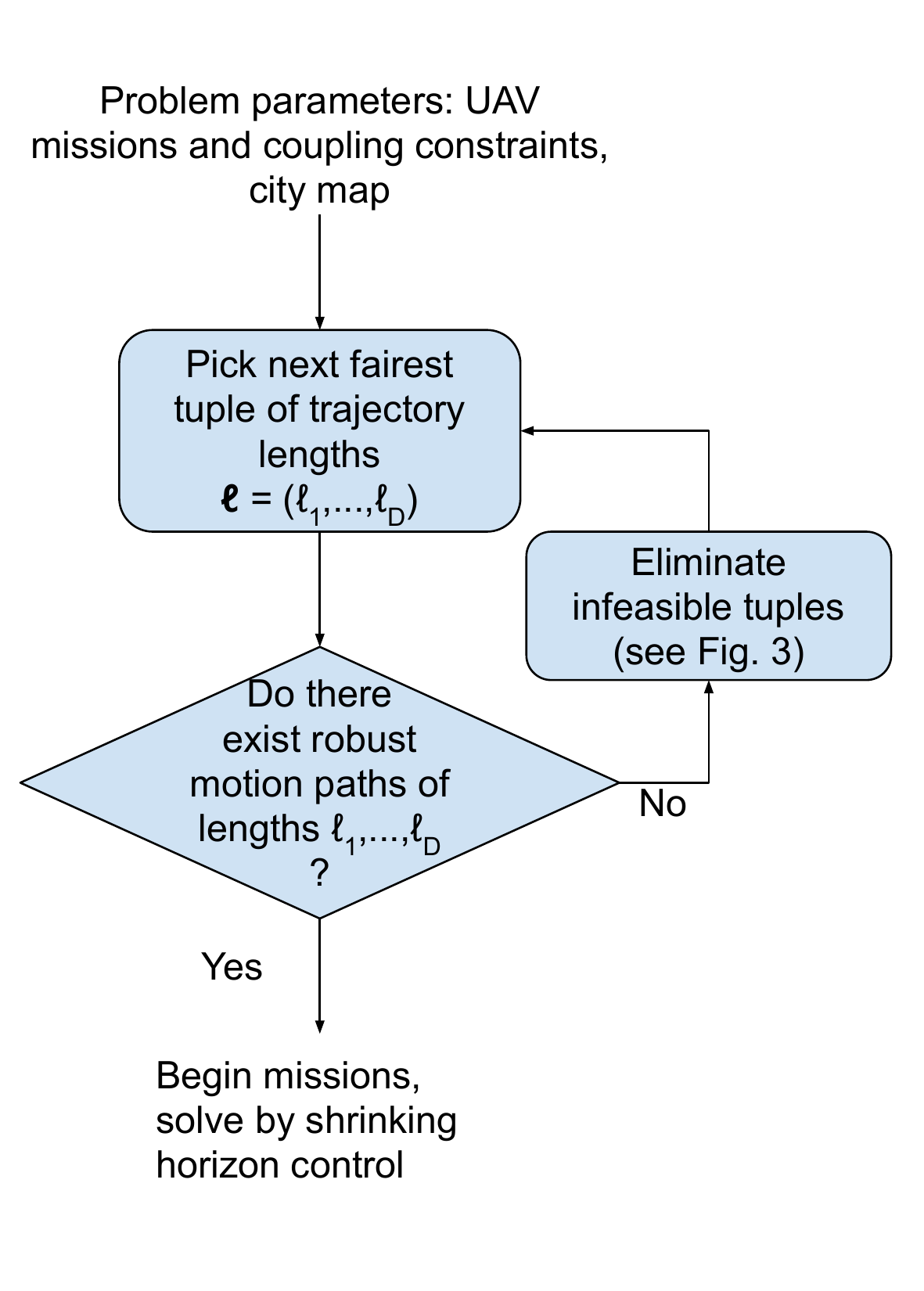}
\smcaption{The \FFbl~algorithm}
\label{fig:FFbl}
\end{figure}

\textbf{Missions formalization}.
In our approach, we formalize the complex missions of Urban Air Mobility (UAM) applications as formulas in \textit{Signal Temporal Logic} (STL)~\cite{Donze10STLRob}.
STL can be thought of as Boolean logic with added temporal operators to capture temporal behavior.
It allows the succinct and unambiguous specification of a wide variety of complex system behaviors over time~\cite{AbbasF13acc,Fainekos18xValuedLTL,Dimarogonas19xagent} and has been used extensively to formalize control objectives, e.g.~\cite{Dimarogonas19xagent}. 
Due to space limitations we refer the reader to~\cite{Donze10STLRob} for formal semantics; we introduce STL via examples.
For example, the specification ``UAV1 reaches the Park within 10 time steps and avoids obstacles on the way there'' is formalized as
\[\formula_1  =  \eventually_{[0,10]}(x_1 \in P) \land \always_{[0,10]} (x_1\notin Obs)\]
in which $\eventually$ is the Eventually operator, $\always$ is Always, and $\land$ is Boolean AND.
Now suppose there is another UAV in the airspace with mission ``UAV2 reaches the bridge within 5 steps and avoids obstacles'':
\[\formula_2 =  \eventually_{[0,5]}(x_2 \in B)  \land \always_{[0,5]} (x_2 \notin Obs)\]
Although the UAVs are independently operated, they do share the airspace, so we must add a mutual separation formula:
\[\formula_3^c = \always_{[0,10]} (\|x_1-x_2\|>s) \]
in which $s$ is a lower bound on the inter-drone separation, set by regulators for example.
We call $\formula_3$ a \textit{coupling constraint}.
In general, a coupling constraint is an STL formula that creates a dependency between the behaviors of 2 or more UAVs.
We say the \textit{global mission} is then
\[\formula_g = \formula_1 \land \formula_2 \land \formula_3^c \]
The specification ``UAV2 stays out of Zone 1 \textit{until} UAV3 exits it, which happens in the next 5 steps'' is formalized as
\[\formula_4 = (x_2 \notin Z_1) \until_{[0,5]} (x_3 \notin Z_1)\]
in which $\until$ is the Until operator.
Interfaces for visualizing formulas~\cite{Hoxha14towardsformal} and specifying missions~\cite{pant2018fly} have been created.

\begin{definition}
	\label{def:global formula}
	 Given $D$ UAVs with their respective missions $\formula_n, 1\leq n\leq D$ and $N$ coupling constraints $\formula^c_n,1\leq n\leq N$, the \textit{global fleet mission} is 
\begin{equation}
\label{eq:global-mission}
\formula_g = \formula_1 \land \ldots \land \formula_D \land  \formula^c_1\land \ldots \land \formula^c_N
\end{equation}
\end{definition}

\textbf{Formula horizon.} All missions/formulas we work with have a finite \textit{horizon}, i.e., they can be satisfied by a finite-length trajectory $\sttraj^H$. 
The horizon $H \in \Ne$ can be calculated from $\formula$ directly~\cite{Dokhanchi14_OnlineMonitoring}, and will be denoted $hrz(\formula)$.
For instance, $\formula_1$ above has a horizon of 10+1=11 (since we count from 0), and $\formula_2$'s horizon is 6.
Two important remarks are in order:
\\
R1) The horizon $H$ is an \textit{upper bound} on the length of a satisfying trajectory: \yhl{if all trajectories of length $H$ violate the formula, then there are no satisfying trajectories of any length.}
A shorter satisfying trajectory might exist.
For instance, $\formula_2$ has horizon 6, but a length-2 trajectory in which UAV2 reaches its goal at $k=1$ does satisfy the formula, and indeed is preferable because it's more efficient. 
\\
R2) The horizon of the global formula $\formula_g$ is greater than any one missions's horizon.
This can be easily deduced from the definition of $hrz(\cdot)$ function in~\cite{Dokhanchi14_OnlineMonitoring}.

\textbf{Traditional control problem.}
We must first present the `traditional' control problem and how Fly-by-Logic~\cite{pant2018fly} solves it, before defining the fair version.
Let $\formula_g$ be the global mission~\eqref{eq:global-mission} of the UAV fleet with horizon $H$.
The problem is to compute $D$ sequences of inputs of equal length, $\inpSig_1^{H-1},\ldots,\inpSig_D^{H-1}$, one per UAV, such that the resulting trajectory $\sttraj(\inpSig^{H-1})$ satisfies $\formula_g$.
\Fbl~finds these input sequences in a centralized fashion by maximizing the \textit{robustness} function $\robfg$ over all possible input sequences:
\begin{equation}
\label{eq:max rob}
\max_{\inpSig^{H-1}} \robf(\sttraj(\inpSig^{H-1}))
\end{equation}
where $\inpSig^{H-1}=(\inpSig_1^{H-1},\ldots,\inpSig_D^{H-1})$.
It is shown in~\cite{FainekosP09tcs} that a positive maximum implies that the corresponding trajectory $\sttraj(\inpSig^{H-1})$ satisfies $\formula_g$.
See~\cite{pant2018fly} for details about robustness and the maximization problem.

We immediately note that \Fbl~only searches over length-$H$ trajectories (equivalently, length-$(H-1)$ input sequences), 
even though their own missions may have shorter horizons by remark R2. 
And as noted above in R1, 
this also means that UAVs are potentially forced to fly longer,  and solve larger control problems online, than they strictly need to.
%


\section{Fair Control Problem and Solution}
\label{sec:fair ctrl}
\subsection{Problem definition}
\label{sec:fair problem defn}
Consider $D$ trajectories $\sttraj^{\slength_1},\ldots, \sttraj^{\slength_D}$, one per UAV, such that collectively they satisfy the global mission $\formula_g$.
We understand fairness of the trajectories $\{\sttraj^{\slength_n}\}_{n=1}^{D}$ as being a notion defined only on the lengths tuple $\langle \slength_1,\ldots,\slength_D\rangle \defeq \vlength$: this is about how early or late each UAV completes its mission, not about how they fly to do so. (Trajectory length can also serve as proxy for energy consumption).
For instance, consider a 3-UAV fleet with $hrz(\formula_g)$=10, and two possible length tuples:
$\vlength = \langle 8,8,8 \rangle$ and $\vlength' = \langle 2, 9, 10\rangle$.
We want to say that $\vlength'$ is less fair than $\vlength$, because it lets UAV1 finish early and forces UAVs 2 and 3 to finish at the limit of what's possible.
We can perform this reasoning purely by looking at the lengths.
Thus, fairness will be a function $f$ that maps a length tuple $\vlength$ to a real number such that a larger $f$-value implies greater fairness. 

It thus emerges that we need a way to determine \textit{which} length tuples to consider and rank by fairness.
For instance, we can now see that \Fbl~only considers $\langle hrz(\formula_g),\ldots,hrz(\formula_g) \rangle$.
But as noted in R1, shorter satisfying trajectories may be possible.
Thus, we need to build a set $PL(\formula)$ of \textit{promising lengths}.
\begin{definition}
	\label{def:PL}
	i) Let $\formula$ be a one-UAV formula. Then the \textit{promising lengths set} $PL(\formula)$ of $\formula$ is the set of integer lengths $\slength$ such that there exists a sequence $\sttraj^\slength \in X^\slength$ that satisfies $\formula$.
	
	ii) Let $\formula_g$ be a $D$-UAVs formula. Then the \textit{promising lengths set} $PL(\formula_g)$ of $\formula_g$ is the set of $D$-tuples $\vlength = \langle \slength_1,\ldots, \slength_D\rangle$ such that there exist $D$ sequences $\{\sttraj_n^{\slength_n}\}_{n=1}^D$, $\sttraj_n^{\slength_n} \in X^{\slength_n}$, which collectively satisfy $\formula_g$.
\end{definition}
Note that whether the sequence $\sttraj^\slength$ can be flown by the UAV - i.e., whether it's a system trajectory or not - remains to be determined. 
\begin{proposition}
	\label{prop:apx PL}
	It is possible to compute an over-approximation of $PL(\formula)$ using a recursion on the structure of $\formula$. \hfill\IEEEQED
\end{proposition}
Thus, because computing $PL$ only requires knowledge of the formula, it can be computed \textit{offline} by the central planner.
All proofs in this paper are  omitted in the interest of space.

We now define the \textit{Fair Temporal Logic Control Problem}:
\begin{problem}
	\label{prob:fair ctrl}
	Consider a global formula $\formula_g$ over $D$ UAVs, a fairness function $f$, and an initial state $x_0 \in X_0$. 
	The fair control problem is 
\begin{subequations}\label{eq:fair ctrl}
	\begin{eqnarray}
	\max_{\vlength \in PL(\formula_g)}&& f(\vlength) \label{eq:outer opt}
	\\
	\text{s.t.}&& \max_{\inpSig^{\vlength}} \rob_{\formula_g}(\sttraj(\inpSig^{\vlength})) \geq 0 \label{eq:inner opt}
	\\
	&&\quad \text{s.t. } x(0) = x_0 \nonumber
	\end{eqnarray}
\end{subequations}
$\inpSig^{\vlength}$ is short for the set of input sequences $\{\inpSig_n^{\slength_n}\}_{n=1}^D$.
\end{problem}
This is a bi-level optimization: the constraints require solving an \textit{inner optimization} \eqref{eq:inner opt} dependent on the primary decision variable $\vlength$. \yhl{The bi-level optimization is only necessary to be solved in an offline phase.}
We will say that $\vlength$ is \textit{feasible} if the corresponding inner optimization has positive solution.
Thus by solving \eqref{eq:fair ctrl} we seek the fairest feasible tuple of trajectory lengths.
We show how to solve \eqref{eq:fair ctrl} in Section~\ref{sec:solving fair ctrl}.

\begin{remark}
	\label{rem:inner opt dimension}
The dimension of the inner optimization, and therefore the time to solve it, is proportional to $\sum_n \slength_n$.
For \Fbl, $\slength_n = hrz(\formula_g)$ for all $n$, so the dimension is proportional to $D\cdot hrz(\formula_g)$.
\end{remark}
 
\subsection{Fairness functions}
\label{sec:fairness fnts}
How to measure fairness? 
We recognize that there is no `best' notion of fairness, and the results from any choice should be interpreted in light of the application.
We present three fairness function candidates;
our framework can accommodate other application-appropriate functions.

Intuitively, every UAV has a range of promising lengths, which is determined by its own mission but also by the coupling constraints that tie it to other UAVs. 
It is preferable for a UAV to be as close as possible to the lower end of this range, but this might force other UAVs towards the upper ends of their ranges.
\textit{Thus, for fairness, all UAV trajectory lengths will lie roughly around the same point in their respective ranges.}
This intuition is formalized as follows.
Given a $D$-tuple $\vlength \in PL(\formula_g)$, we write $\vlength(n)$ for the $n^{th}$ element of $\vlength$. 
Define 
\[\lhn{n} = \min\{\vlength(n) \such \vlength \in PL(\formula_g)\},~ \uhn{n} = \max\{\vlength(n) \such \vlength \in PL(\formula_g) \}\]
\[\alpha_n = (\slength_n - \lhn{n})/(\uhn{n} - \lhn{n})\]
So $\{\lhn{n},\ldots,\uhn{n}\}$ is the range of lengths that $\vlength(n)$ can take in $PL(\formula_g)$,
and $\alpha_n$ measures the fraction of that range at which a given $\slength_n$ lies.
Note that $\alpha_n$ is a monotone function of $\slength_n$.
Given $\vlength$ and the corresponding tuple of fractions $\vec{\alpha} = \langle \alpha_1,\ldots,\alpha_n\rangle$, the first fairness function is simply the negative of the variance of the $\alpha_n$'s:
\begin{equation}
\label{eq:f1}
f_1(\vlength) = - \var(\vec{\alpha})
\end{equation}
(we use negative variance because we want to maximize fairness).

$f_1$ is a negative function with maximum 0, achieved when all $\alpha_n$'s are equal, i.e. when all UAVs trajectory lengths are exactly at the same fraction in their length ranges $\{\lhn{n}, \uhn{n}\}$.

Now all else being equal, a solution where every $\alpha_n = 1/2$, say, is preferable to a solution where every $\alpha_n = 1$, since a smaller $\alpha$ means a shorter and more efficient trajectory. 
Moreover, if a (3-UAV) solution with $\vec{\alpha} = \langle 3/4,3/4,1/2 \rangle$ is feasible, forcing a solution with $\vec{\alpha} = \langle 3/4, 3/4,1 \rangle$ seems unfair - after all, a shorter trajectory for the 3rd UAV is not forcing longer trajectories for the others.
The second fairness function captures this by adding a regularizer to $f_1$ which encourages small lengths, thus balancing between balance and efficiency:
\begin{equation}
\label{eq:f2}
f_2(\vlength;w) = - w\var(\vec{\alpha}) - (1-w)\sum_{n=1}^{D}\alpha_n^2 
\end{equation}
Here, $w \in (0,1]$ is a weighting factor.

The last fairness function allows for explicit negotiation between operators for preferential treatment: this is a legitimate use-case, in which some operators pay to choose trajectory lengths that suit them, unconstrained by fairness considerations. 
Given a weight tuple $\vec{v} \in (0,\infty)^D$, define
\begin{equation}
\label{eq:f2imb}
f_2^{imb}(\vlength;w,\vec{v}) = -w\var(\vec{\alpha}) - (1-w)\sum_{n=1}^{D}v_n\alpha_n^2 
\end{equation}
As $v_n$ grows larger,  $\alpha_n$ gets valued more in the fairness weighting, thus favoring UAV$_n$.

\subsection{Solving the Fair Control Problem}
\label{sec:solving fair ctrl}
The control problem is solved in an offline and online phases.
Offline, we find one solution to \eqref{eq:fair ctrl}, which is a promising length tuple $\vlength^* = \langle \slength_1^*,\ldots, \slength_n^*\rangle$ and satisfying trajectory $\sttraj^*$ in which UAV$_n$ has a trajectory of length $\slength_n^*$.
Online, i.e. after take-off, a classical shrinking horizon procedure is implemented to continuously update the trajectory $\sttraj^*$ based on the latest state estimate.
See~\cite{pant2018fly} for details.
Because the online phase is a special case of the offline phase, we focus on the latter.

The offline phase could be solved as follows.
See Fig.~\ref{fig:FFbl}.
The set $PL(\formula_g)$ is finite: 
 start with a fairest promising length tuple (one which maximizes $f(\vlength)$) and check whether it is feasible - i.e., whether the corresponding inner maximization has positive maximum.
If yes, we are done. 
Else, we pick the next fairest promising tuple $\vlength$ and repeat, until we find the fairest promising length whose inner optimization has positive \yhl{robustness using \Fbl}.

\begin{figure}[t]
\centering
\includegraphics[width=0.7\linewidth]{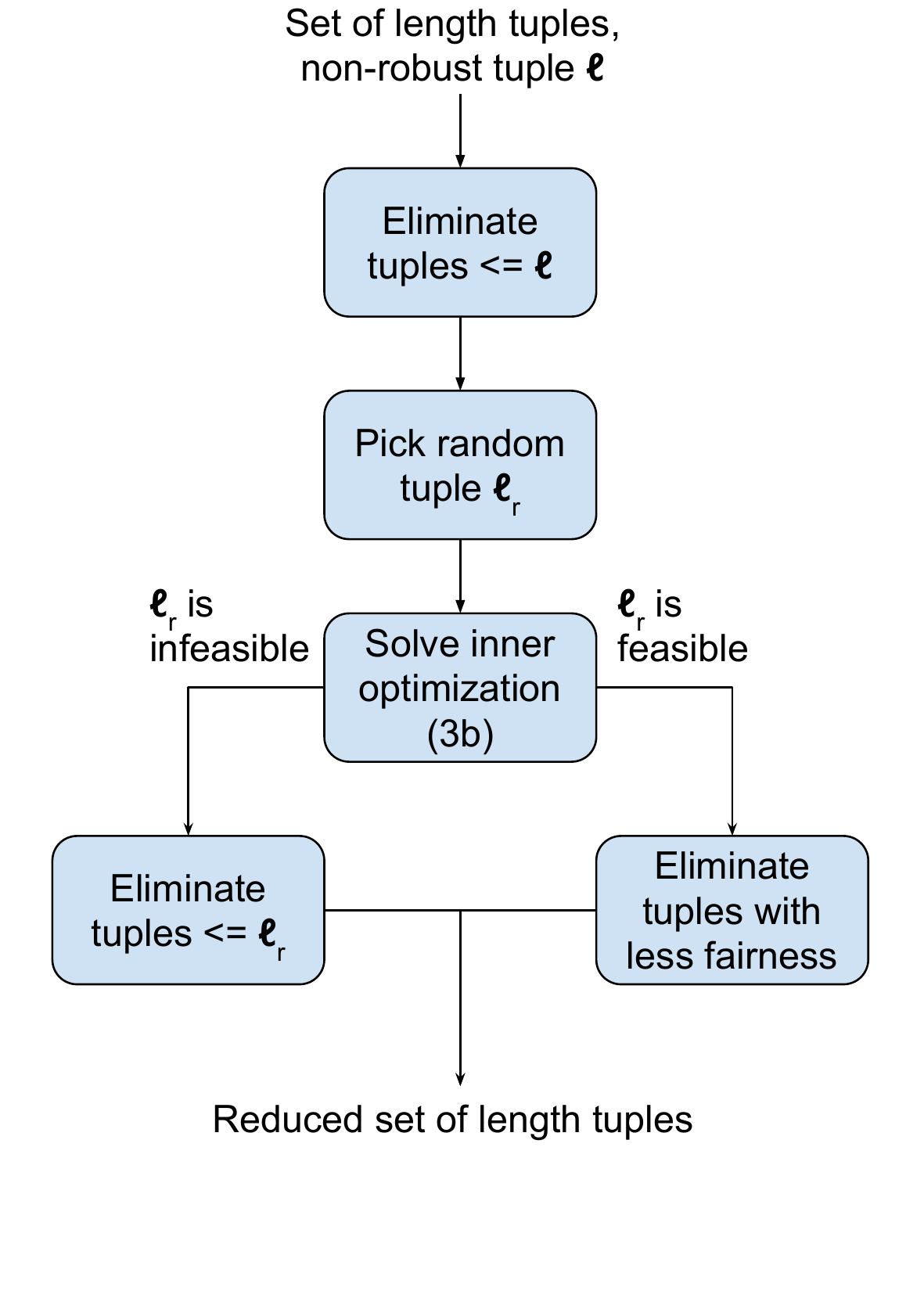}
\smcaption{Sound reduction of search space for outer optimization~\eqref{eq:outer opt}. This is step ``Eliminate infeasible tuples'' in Fig.~\ref{fig:FFbl}.}
\label{fig:space reduction}
\end{figure}

This brute force approach is only possible for small numbers of UAVs and small horizons since the size of $PL$ grows as $O(H^D)$.
For larger $D$ or $H$ it is not even possible to store $PL$ in memory. 
Therefore, in our implementation of \FFbl, we don't build $PL$ in memory, we use instead an implicit representation.
The search is made more efficient by the following proposition.
\begin{proposition}
	\label{prop:binary search}
	(a) If $\vlength$ is  infeasible, then every smaller $\vlength$ in lexicographic order is also infeasible. 
	\\
	(b) If $\vlength$ is feasible, then the optimizer of \eqref{eq:fair ctrl} has fairness at least $f(\vlength)$.\hfill\IEEEQED
\end{proposition}
This proposition allows us to reduce the search space of \eqref{eq:outer opt} with every iteration.
See Fig.~\ref{fig:space reduction}: 
Prop.~\ref{prop:binary search}(a) says that every infeasible promising tuple allows us to eliminate from consideration all smaller tuples in one go.
Therefore we store every infeasible tuple we encounter to check whether future tuples are smaller than it; if yes, we skip them without wasting time solving the inner optimization for them, which is the real computational bottleneck.

Prop.~\ref{prop:binary search}(b) allows us to eliminate tuples that are less fair than feasible tuples we encounter.
Therefore, we occasionally randomly sample the set $PL$ as shown in Fig.~\ref{fig:space reduction}: if $\vlength_r$ is feasible, we store it and compare future tuples to it. If they are less fair, we skip them. 
Note that as soon as an elimination takes place the remaining search space becomes non-convex (as a compact subset of $\Re^D$). 
Therefore, the step to pick the next fairest tuple in Fig.~\ref{fig:FFbl} 
only yields local optima. 

\section{Experiments}
\label{sec:exp}
\newcommand{\ninits}{20}
We implemented our solution, called \FFbl, on top of Fly-by-Logic, a toolbox for motion planning and control of quadrotor fleets~\cite{pant2017smooth,pant2018fly}. \yhl{This is implemented by expanding \Fbl~to solve using distinct horizons for each UAV instead of a global horizon, and then implementing the outer optimization.}
We compare the solutions provided by default \Fbl~(without fairness considerations) and \FFbl.
All simulations were run with an Intel CPU at 2.60 GHz on a single core. 

\subsection{The effects of fairness}
\label{eq:exp fairness}
We report the results of quadrotor fleet simulations for Reach-Avoid missions.
The Reach-Avoid formula for $D$ quadrotors is 
\begin{equation}
\label{eq:reach-avoid}
\begin{aligned}[m]
\formula &= \formula_{G}\land\formula_{O}\land\formula_{M}  \\
\formula_{G} &= \bigwedge_{n=1}^{D}\eventually_{[0,H_n]}(x_n \in G_n) \\
\formula_{O} &= \bigwedge_{n=1}^{D}\always_{[0,H_n]}(x_n \notin O) \\
\formula_{M} &= \bigwedge_{n,m=1, n\neq m}^{D}\always_{[0,H_{n,m}]}(\|x_n-x_m\| > d) \\
\end{aligned}
\end{equation}
in which $H_n$ is the horizon of UAV$_n$'s mission, $G_n$ is the goal of UAV$_n$, $O$ is the set of obstacles that all UAVs are avoiding, $d$ is the minimum inter-drone distance, and $H_{n,m}=\min(H_n,H_m)$ (UAV$_n$ and UAV$_m$ need to avoid each other only for as long as both are flying). \yhl{$\bigwedge_{n=1}^{D}\phi_n$ represents the conjunction of every $\phi_n$ for $1\leq n\leq D$}

We ran experiments with $D=5,10,15$ and $20$ quadrotors.
For each value of $D$, we ran three algorithms to solve \eqref{eq:fair ctrl}: default \Fbl~and \FFbl~using both $f_1$ and $f_2$. For each algorithm, and to have meaningful results, we solve the fair control problem \eqref{eq:fair ctrl} \ninits~times, each time starting from a different initial state $\stPt_0$. \yhl{The same map is used for each iteration of the experiment. The map is left simple with a single obstacle that the UAVs must avoid, and the goals of each UAV are held constant.}

We compared: 
	a) the average robustness of the three solutions, 
	b) the average fairness of the three solutions, 
	c) the average time is takes to solve the offline phase, and 
	d) the average time to complete the first iteration of the online control phase (subsequent iterations take less time).
All averages are over the \ninits~initial states.


\begin{figure}[t]
	\centering
	\includegraphics[width=0.7\linewidth]{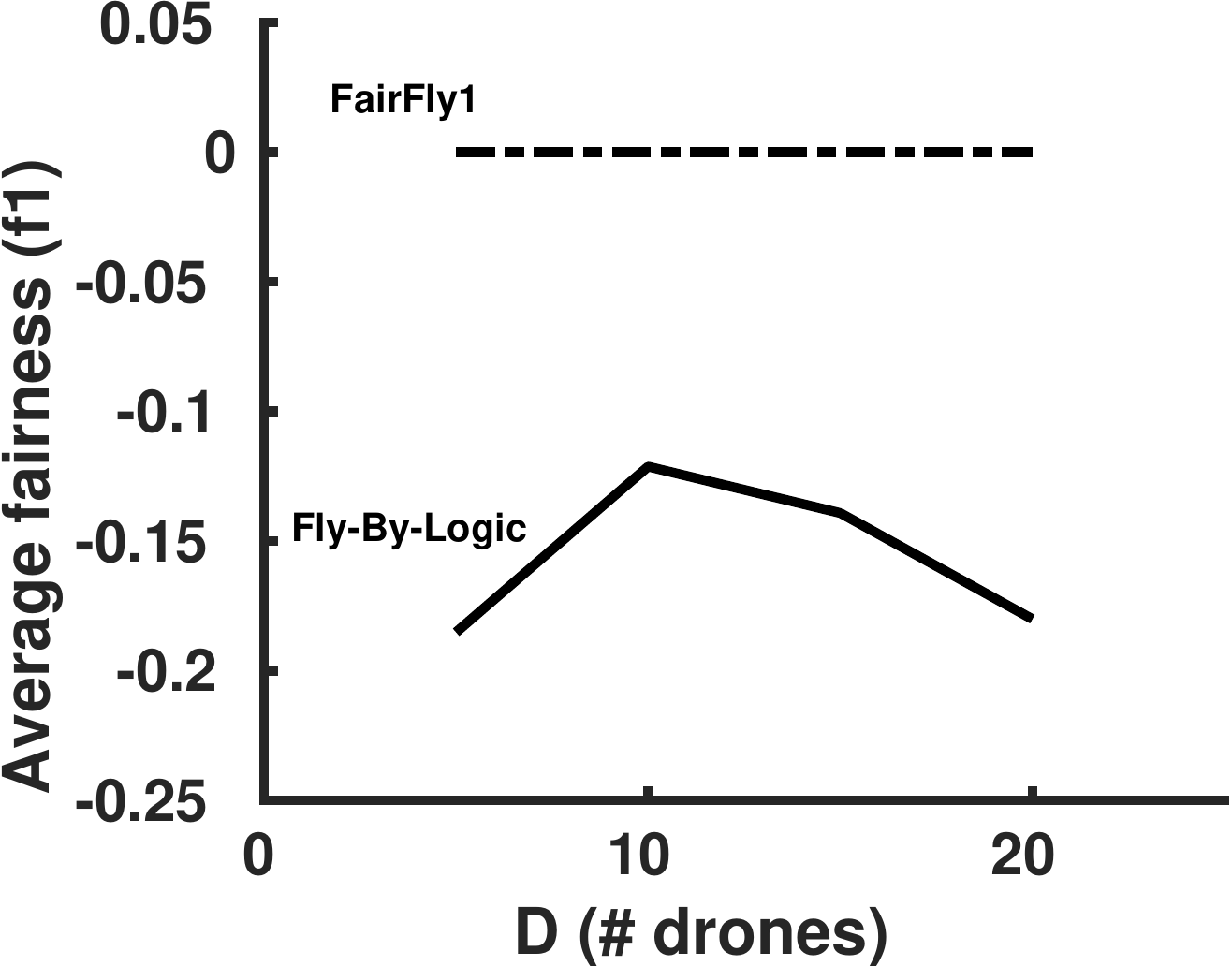}
	\smcaption{Average fairness of solutions found by \Fbl~(solid) and \FFbl~(dash dotted) using $f_1$.}
	\label{fig:fairness comparison1}
\end{figure}

\begin{figure}[t]
	\centering
	\includegraphics[width=0.7\linewidth]{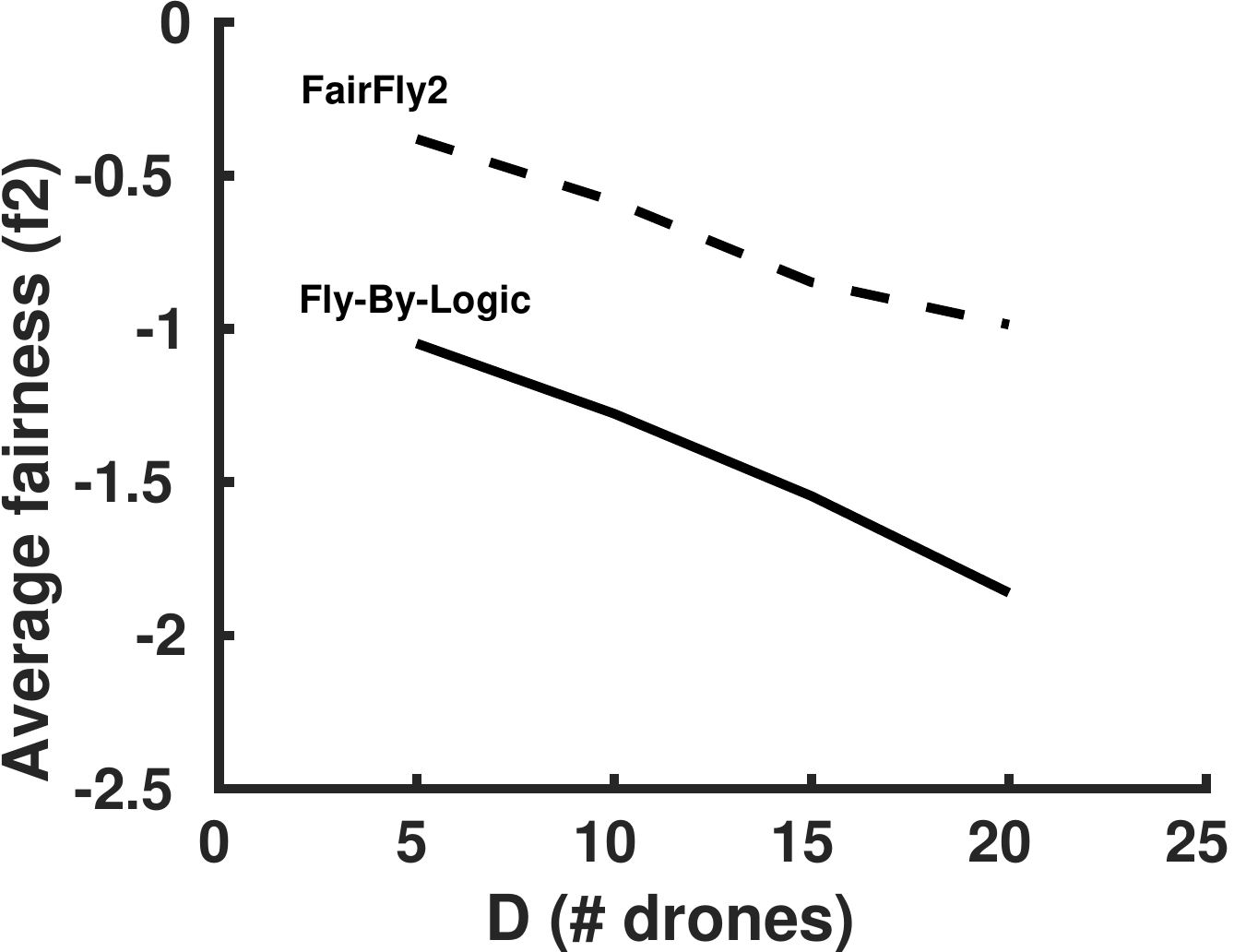}
	\smcaption{Average fairness of solutions found by \Fbl~(solid) and \FFbl~(dashed) using $f_2$.}
	\label{fig:fairness comparison2}
\end{figure}

\textbf{Fairness Results.}
See Figs.~\ref{fig:fairness comparison1} and~\ref{fig:fairness comparison2}.
Using either fairness notion, the solutions obtained by \FFbl~are more fair than ones that \Fbl~outputs for any number of UAVs. With $f_1$, the fairest solution is when all $\alpha$'s are equal. The maximizer that gets chosen by \FFbl~is when all $\alpha$'s are equal to 1, or equivalently $\vlength=\langle H_1, H_2, ... H_D\rangle$. 
Note that if this length tuple is not feasible, then no other tuple can be, so only a maximum fairness solution needs to be checked.

With $f_2$, the fairness value declines as more UAVs are added. This is expected: with more drones the airspace is more densely occupied, requiring more compromises to the trajectories of the quadrotors, thus increasing the impact of the efficiency factor in $f_2$.

\begin{figure}[t]
	\centering
	\includegraphics[width=0.7\linewidth]{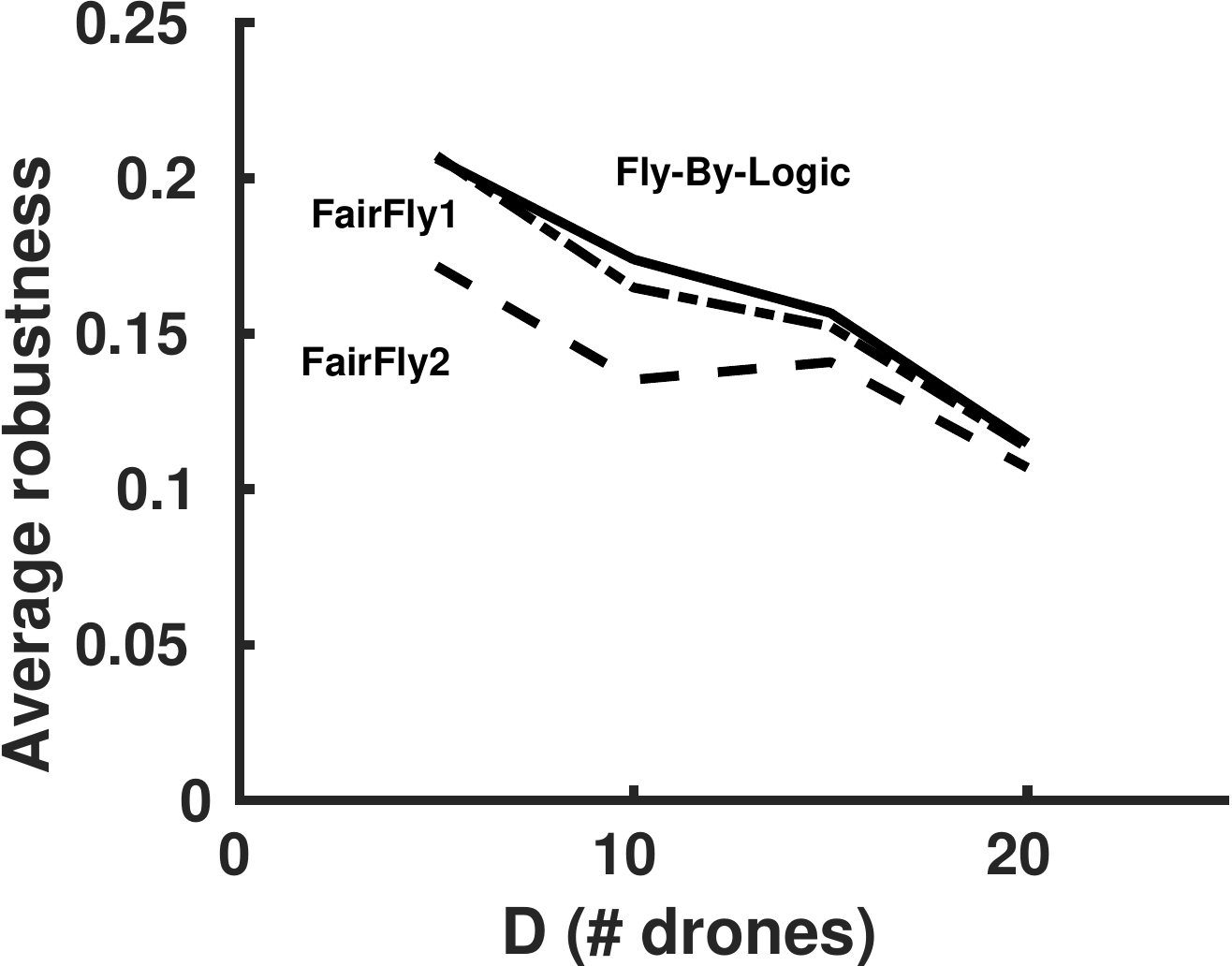}
	\smcaption{Average robustness of solutions found by \Fbl~(solid) and \FFbl~with $f_1$ (dash dotted), and $f_2$ (dashed)}
	\label{fig:robustness}
\end{figure}

\textbf{Effects on robustness.}
Fig.~\ref{fig:robustness} shows the robustness of the optimal trajectories returned by \Fbl~and \FFbl~using both $f_1$ and $f_2$.
Using $f_1$ is slightly less robust on average than \Fbl, and $f_2$ is less robust than $f_1$. 
It is noted that the reduction in robustness is not dramatic, and more importantly, robust trajectories still exist while increasing fairness.

\begin{figure}[t]
	\centering
	\includegraphics[width=0.7\linewidth]{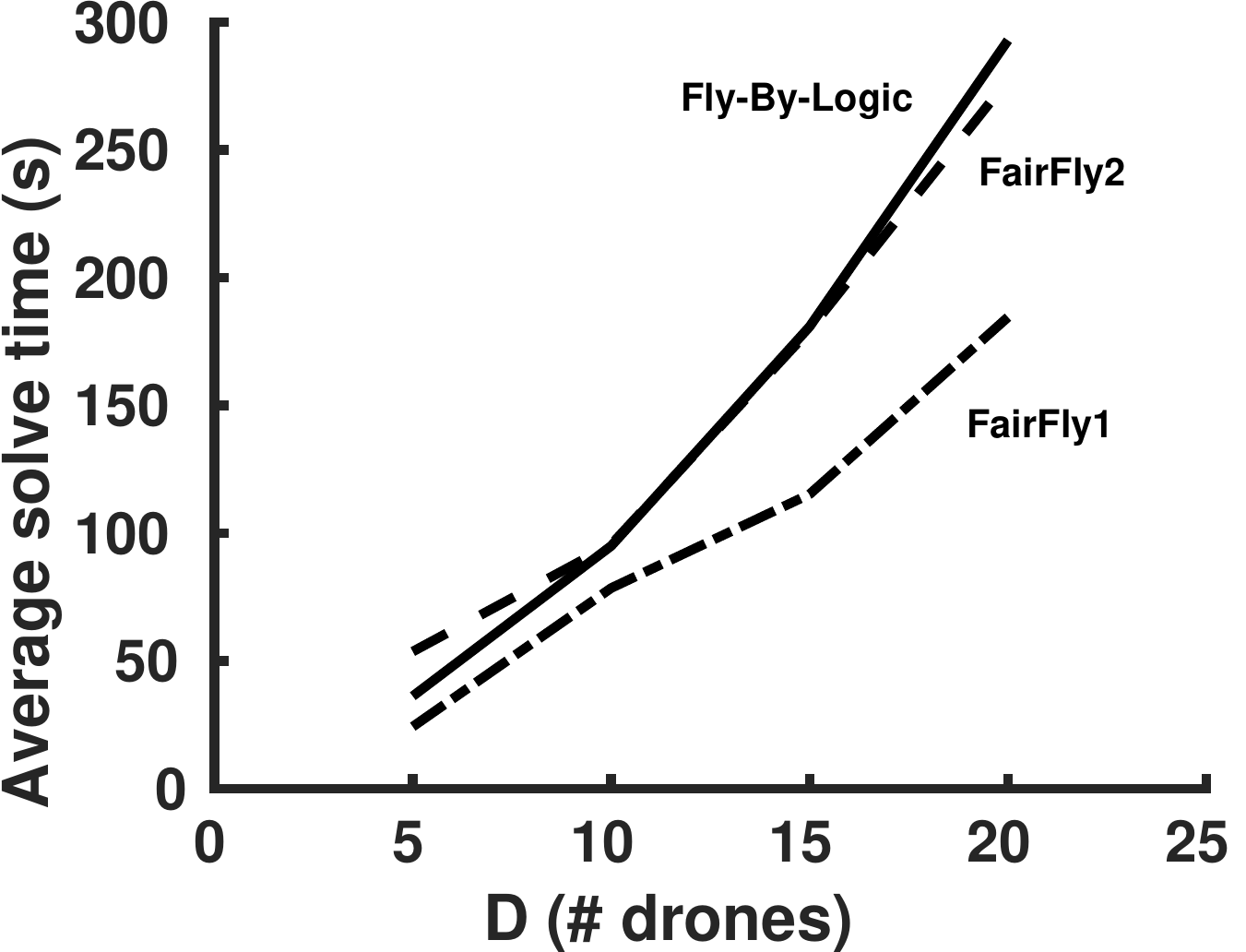}
	\smcaption{Time to solve the offline optimization for \Fbl~(solid) and \FFbl~with $f_1$ (dash dotted), and $f_2$ (dashed), averaged over choice of initial states.}
	\label{fig:offline solve}
\end{figure}

\textbf{Offline computational overhead.}
Now we examine the computational overhead of finding fair trajectories. 
See Fig.~\ref{fig:offline solve}.
\FFbl~with $f_1$ is actually quicker offline than \Fbl. 
The explanation is this: as noted above, an $f_1$-fairest trajectory tuple is equal to the individual horizon of each UAV, i.e. $\vlength=\langle H_1,\ldots,H_D\rangle$.
If this tuple is not feasible, than no other tuple can be feasible, so only this one length tuple needs to be checked for feasibility. 
By Remark~\ref{rem:inner opt dimension}, the size of the inner optimization for \FFbl~is smaller than that of \Fbl, so it is quicker to solve.

\FFbl~calculated with $f_2$ was slightly slower than \Fbl. 
This is while acknowledging the fact that \FFbl was not necessarily producing the globally $f_2$-fairest result because it is using a non-convex optimization to find the next fairest length tuple. 

\begin{figure}[t]
	\centering
	\includegraphics[width=0.7\linewidth]{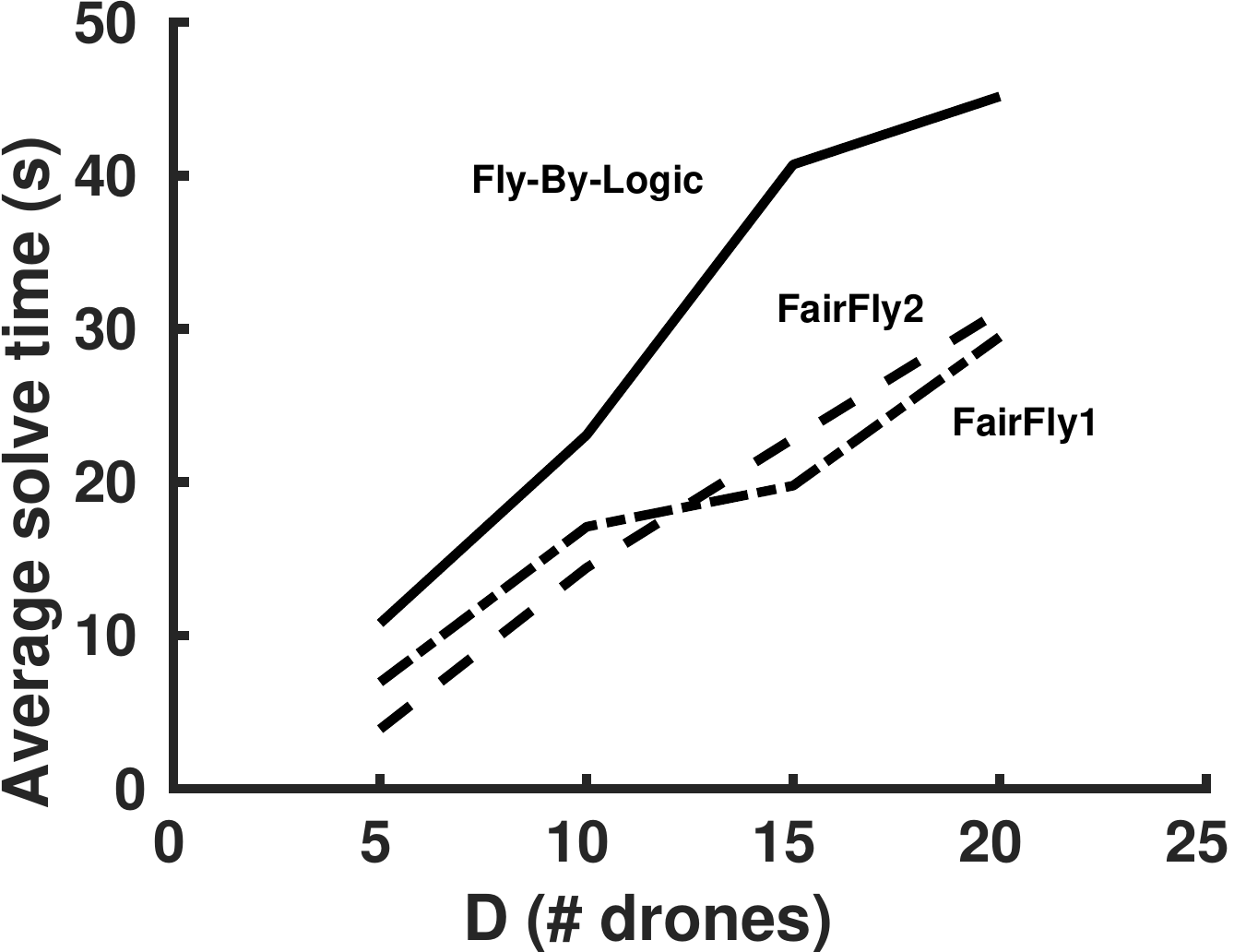}
	\smcaption{Time to solve first iteration of the online control problem in shrinking horizon fashion for \Fbl~(solid) and \FFbl~with $f_1$ (dash dotted), and $f_2$ (dashed), averaged over choice of initial states.}
	\label{fig:online solve}
\end{figure}



\textbf{Online computational gains.}
We now study the online run time performance of \FFbl, as this will directly impact online control stability of the UAVs. 
The offline phase yields a fairest feasible length tuple $\vlength^*$ and a corresponding most robust trajectory $\sttraj^*$. 
In the first iteration of the online phase 
the first control action from $\sstraj^*$ is applied.
At the second iteration, the inner maximization is re-solved with one less time-step (since we already did one step).
More generally, at the $k^{th}$ iteration, the inner maximization is re-solved with $k$ fewer time-steps (since we already did $k$ steps).
Therefore, the horizons that the optimization must solve for shrink every time step.

For our analysis we simply looked at the run time for the first iteration of the online phase.
See Fig.~\ref{fig:online solve}. 
The \FFbl~variations greatly distinguish themselves from \Fbl~here. The run time of \FFbl~is \yhl{about} 50\% faster than \Fbl~with $f_2$ being fastest on average. This is expected; $f_2$ is solving for the shortest overall trajectories due to the length regularizer term (see Fig.~\ref{eq:f2}). 
Because the length of trajectories is shorter in $f_2$, the overall search space is reduced for the inner optimization.

\subsection{Comparing the fairness functions}
\label{sec:exp comparing fairness}
\yhl{Now we will look at the length trajectories that each fairness function generates, in order to conceptualize what each function considers as "fairest".}

To compare the fairness functions $f_1$ and $f_2$, we ran a $D=5$ Reach-Avoid experiment with individual horizons $H_1=10, H_2=8, H_3=5, H_4=6, H_5=7$ (see~\eqref{eq:reach-avoid}).
With $f_1$, the fairest length tuple is $\vlength=\langle10, 8, 5, 6, 7\rangle$.
With $f_2(\cdot; w=0.75)$  the fairest length tuple is $\vlength=\langle9, 7, 5, 6, 6\rangle$.
It is noted that with $f_2$ every UAV finishes at least as fast as it would have with $f_1$, and some can finish quicker. With $f_1$, every UAV travels a mission that is the longest allowed by their horizon.
No UAV is allowed to complete their mission any quicker, but neither are any of the UAVs required to continue flying until the rest finish their mission.

\yhl{We then ran a case where UAVs negotiate an imbalanced solution to compare our third notion of fairness, as shown in~\eqref{eq:f2imb}.}  
We ran the same $D=5$ experiment as above with $f_{2}^{imb}$ with $v_1=v_2=10$, and $v_3=v_4=v_5=1$. 
This gives advantage to the first two quadrotors at the expense of the last three. 
The fairest length vector is $\vlength=\langle7, 6, 5, 6, 7\rangle$, compared to the solution from $f_2$, namely $\langle 9, 7, 5, 6, 6 \rangle$. As expected, the first two UAVs received shorter trajectories, while the last three had the same or longer trajectories. 

\section{Conclusions}
\label{sec:conclusions}
\FFbl~was shown to produce both fairer and more efficient trajectories for UAVs with a slight impact to trajectory robustness. 
Future work will focus on faster offline optimization and hardware implementations.







\bibliographystyle{IEEEtran}
\bibliography{iccps2017,hscc17,hscc2016,hscc19,cav2019,fainekos_bibrefs,itsc2020}

\end{document}